\documentclass[aps,preprint,epsfig,rotate]{revtex4}
\usepackage{graphicx}
\usepackage{bm}
\usepackage{epsfig}

\input epsf
\begin{document}
\title{Matrix mechanics for actual atoms and molecules}

\author{Alexei M. Frolov}
\email[E--mail address: ]{alex1975frol@gmail.com}


\affiliation{Department of Applied Mathematics \\
       University of Western Ontario, London, Ontario N6H 5B7, Canada} 

\date{\today}

\begin{abstract}

Matrix mechanics is developed to describe the bound state spectra in few- and many-electron atoms, ions and molecules. Our method is based on the matrix 
factorization of many-electron (or many-particle) Coulomb Hamiltonians which are written in hyperspherical coordinates. As follows from the results of 
our study the bound state spectra of many-electron (or many-particle) Coulomb Hamiltonians always have the `ladder' structure and this fundamental fact 
can be used to determine and investigate the bound states in various few- and many-body Coulomb systems.    

\noindent 
PACS number(s): 31.15.-A, 31.15.ac and 32.30.-r

\end{abstract}

\maketitle
\newpage

\section{Introduction}

In this communication we develop the matrix mechanics of the actual, i.e. few- and many-electron, atoms, ions and molecular systems. This approach is, in fact, 
a very powerful method for analysis of various few- and many-electron Coulomb systems which can successfully be applied to describe the bound state spectra in 
different atoms, ions and even molecules. Our approach is a new step in the development of Matrix Mechanics \cite{Green} which was originally created by 
Heisenberg, Born and Jordan as the first version of Quantum Mechanics \cite{Heis}, \cite{Heis2}. Briefly, we want to show how the old version of matrix mechanics 
can be modified to the new level and can be used as an effective tool for solving numerous problems in modern atomic physics. 

Our main goal in this study is to show that the Coulomb Hamiltonian of an arbitrary atom which contain $N_e-$bound electrons is always factorized, i.e. it is 
represented in the form of a product of the two differential operators of the first order. This fundamental fact is directly related to the internal structure 
of the Coulomb Hamiltonians (the so-called ladder structure) and substantially simplifies analysis of the bound state spectra in few- and many-electron atoms and 
ions and can be used to perform more accurate numerical computations of the bound states. In particular, by using the method of matrix factorization we can 
determine the energies and wave functions of an arbitrary bound state in many-electron atoms and ions, including excited and highly excited bound states. The same 
procedure can also be used for molecules and for other many-particle Coulomb systems.    

First, let us consider the one-electron hydrogen atom and/or hydrogen-like ions, i.e. atomic systems which contain one bound electron and one positively charged 
nucleus. To simplify our analysis in this study we shall assume that all atomic nuclei mentioned below are infinitely heavy. Furthermore, everywhere below we 
shall apply the atomic system of units. In these units $\hbar = 1, \mid e \mid = 1$ and $m_e = 1$, where $\hbar = \frac{h}{2 \pi}$ is the reduced Planck constant, 
$m_e$ is the electron mass and $e$ is the electric charge of electron (a negative value). In atomic units the Hamiltonian of one-electron atoms/ions is written in 
the form
\begin{eqnarray}
 H = -\frac{\hbar^2}{2 m_e} \Bigl[ \frac{\partial^2}{\partial r^2} + \frac{2}{r} \frac{\partial}{\partial r} - \frac{{\bf L}^2}{r^{2}} \Bigr] - \frac{Q e^2}{r} 
 = -\frac{1}{2} \Bigl[ \frac{\partial^2}{\partial r^2} + \frac{2}{r} \frac{\partial}{\partial r} - \frac{{\bf L}^2}{r^{2}} \Bigr] - \frac{Q}{r} \; \; \; 
 \label{eq1}
\end{eqnarray} 
where $Q e = Q$ is the electric charge of the atomic nucleus and ${\bf L}$ is the operator of the angular moment of the atom which coincides with the total angular 
momentum of the bound atomic electrons. To determine the bound 
states in the hydrogen atom and hydrogen-like ions we need to solve the corresponding Schr\"{o}dinger equation $H \Psi = E \Psi$, where the operator $H$ is the 
Hamiltonian, Eq.(\ref{eq1}), $\Psi$ is the unknown wave function and $E$ is the eigenvalue of $H$ which is the total energy of the bound state, i.e. $E < 0$. As is 
well known (see, e.g., \cite{Dirac}, \cite{LLQ}) the total wave function of an arbitrary bound state of the hydrogen atom is represented as a product of the radial 
part of the total wave function $\psi_{n \ell}(r)$ and the corresponding spherical harmonic(s) $Y_{\ell m}(\theta, \phi)$, i.e. $\Psi_{n \ell m}(r, \theta, \phi) = 
\psi_{n \ell}(r) Y_{\ell m}(\theta, \phi)$, where $Y_{\ell m}(\theta, \phi)$ are the eigenfunctions of the ${\bf L}^2$ operator, i.e. ${\bf L}^2 Y_{\ell m}(\theta, 
\phi) = \vec{\ell}^2 Y_{\ell m}(\theta, \phi) = \ell (\ell + 1) Y_{\ell m}(\theta, \phi)$. Here and everywhere below the notations $\theta$ and $\phi$ stand for the 
spherical coordinates of the bound electron, while $r$ is the electron-proton distance which coincides with the radial spherical coordinate. The integer numbers $n, 
\ell$ and $m$ are called the principal quantum number, angular quantum number and magenetic quantum number, respectively. For one-electron atomic systems all these 
quantum numbers are the `good' (or conserving) quantum numbers. Note also that the following inequalies are always obeyed for these quantum numbers: $\ell \le n - 1$ 
and $\mid m \mid \le \ell$.  

In the basis of spherical harmonics $Y_{\ell m}(\theta, \phi)$ the Hamiltonian, Eq.(\ref{eq1}), takes the form
\begin{eqnarray}
  H(r) = -\frac{1}{2} \Bigl[ \frac{\partial^2}{\partial r^2} + \frac{2}{r} \frac{\partial}{\partial r} \Bigr] + \frac{\ell (\ell + 1)}{2 r^{2}} - \frac{Q}{r} 
 = \delta_{\ell,\ell_{1}} \delta_{m,m_{1}} \Bigl\{-\frac{1}{2} \Bigl[ \frac{\partial^2}{\partial r^2} + \frac{2}{r} \frac{\partial}{\partial r} \Bigr] + 
 \frac{\ell (\ell + 1)}{2 r^{2}} - \frac{Q}{r} \Bigr\} \; \; \; \label{Hamilt}
\end{eqnarray} 
where $\ell$ is the angular moment of the bound electron ($\ell \ge 0$) which coincides (for one-electron atoms/ions) with the angular momentum of the whole atom $L$. 
Note that the Hamiltonian, Eq.(\ref{Hamilt}) is a differential operator of the second order upon the radial variable $r$. On the other hand, the Hamiltonian $H(r)$ is 
a diagonal matrix in terms of the $\ell$ and $m$ (or $\mid m \mid$) indeces each of which is a conserving quantum number. The explicit solution of the Schr\"{o}dinger 
equation $H \psi = E \psi$ for the bound states of the hydrogen atom and hydrogen-like ions leads to the following formula (Borh's formula) for the energy spectrum
\begin{equation}
  E_n = -\frac{m_e Q^2 e^{4}}{2 \hbar^2 n^2} = -\frac{Q^2}{2 n^2} = -\frac{Q^2}{2 (n_r + \ell + 1)^2} \label{spectr1}
\end{equation}
where $n_r$ is the radial quantum number which is a non-negative integer and varies between 0 and $n - \ell - 1$ and $n$ is the principal quantum number. The 
numerical value of $n_r$ coincides with the number of zeros in the radial part of the wave function $\psi_{n\ell}(r)$. Furthermore, the radial part of the total 
wave function $\psi_{n\ell}(r)$ equals to the product of some positive power of $r$, Laguerre polynomial of $r$ and a radial exponent. In numerous textbooks this 
results is derived by using a special form of the radial wave function $\psi_{n \ell}(r)$ (see, e.g., \cite{LLQ}). Then the original differential equation is 
reduced to the corresponding differential equation for the hypergeometric function ${}_2F_{1}(a, b; c; r)$ which must have a finite number of terms, or, in other 
words, to be a polynomial. This is the standard procedure which have been described in many textbooks. However, there is another procedure which can be applied 
to determine the bound state spectrum, i.e. the total energies and wave functions, of the hydrogen atom and hydrogen-like ions. This procedure is more elegant, 
physically transparent and based on the internal structure of the Coulomb Hamiltonian (see, e.g., \cite{Green}). We describe this procedure in the next Section.

\section{Factorization method for one-electron atom/ion}

This Section is intended merely to summarize the central facts about the factorization method that are needed in Sections IV - V below. Another aim 
of this Section is to fix the notation. Now, consider the matrix of the Hamiltonian $H$, Eq.(\ref{eq1}), in the basis of spherical harmonics, i.e. the matrix 
$\langle Y_{\ell m}(\theta, \phi) \mid H \mid Y_{\ell_1 m_1}(\theta, \phi) \rangle = \delta_{\ell, \ell_{1}} \delta_{m, m_{1}} \hat{H}_{\ell, m}(r) = H(r)$ which 
is a diagonal matrix in the $\ell$ and $m$ indices. On the other hand, each matrix element of this matrix is a differential operator of the radial variable $r$, 
i.e. $\hat{H}_{\ell, m}(r)$. Since the both $\ell$ and $m$ quantum numbers are the conserving (or `good') quantum numbers, then we can replace the corresponding 
matrix notation $\hat{H}_{\ell, m}(r)$ by a simple operator notation, i.e., $\hat{H}_{\ell, m}(r) = H(r)$ (see, Eq.(\ref{Hamilt})). Our goal in this Section is to 
find all eigenvalues of this radial operator $H(r)$, Eq.(\ref{Hamilt}). For these purposes we shall apply the factorization method developed for the differential 
operators of the second order. This method was well described in a number of books and textbooks (see, e.g.,  \cite{Green}, \cite{Elut} and references therein). 
Below, we assume that the reader is acquainted with the factorization method and its applications to one-electron atomic systems (see, e.g., \cite{Green} and 
references therein). 

The method of matrix factorization (see, e.g., \cite{Green}) is based on the existence of a set of the first-order differential operators $\Theta_n(r)$ (where 
$n = 1, 2, \ldots$) and their adjoint operators $\Theta^{\ast}_n(r)$. The $\Theta_n(r)$ operators are written in the form
\begin{equation}
 \Theta_n(r) = \frac{1}{\sqrt{2}} \Bigl[ -\Bigl(\frac{\partial}{\partial r} + \frac{1}{r}\Bigr) + \frac{\beta_n}{r} + \alpha_n \Bigl] \; \; \; \label{Thn}
\end{equation}
In respect to this definition the adjoint operators are
\begin{equation}
 \Theta^{\ast}_n(r) = \frac{1}{\sqrt{2}} \Bigl[ \Bigl(\frac{\partial}{\partial r} + \frac{1}{r}\Bigr) + \frac{\beta_n}{r} + \alpha_n \Bigl] \; \; \; 
 \label{Thna}
\end{equation}
The real parameters $\beta_n$ and $\alpha_n$ in operators defined by Eqs.(\ref{Thn}) - (\ref{Thna}) must be chosen to obey the two fundamental conditions of the
factorization method. First, the Hamiltonian $H(r)$, Eq.(\ref{Hamilt}), must be represented in the form
\begin{eqnarray}
  H = \Theta^{\ast}_1(r) \Theta_1(r) + a_1 \; \; \; \label{factH1}
\end{eqnarray} 
where $H$ is the Coulomb Hamiltonian, Eq.(\ref{Hamilt}), of the one-electron hydrogen atom. Second, there is an infinite, in principle, chain of relations 
between the $\Theta_n(r), \Theta^{\ast}_n(r), \Theta^{\ast}_{n+1}(r)$ and $\Theta_{n+1}(r)$ operators:
\begin{eqnarray}
  \Theta_n(r) \Theta^{\ast}_n(r) + a_n = H_{n+1} = \Theta^{\ast}_{n+1}(r) \Theta_{n+1}(r) + a_{n+1} \; \; \; \label{factHn1}
\end{eqnarray}
where $H_{n+1}$ is the $n-$excited Hamiltonian (or $n$-times excited Hamiltonian, where $n \ge 1$) of the original problem. In this notation we have to assume 
that $H_{1} = H$. The equations, Eqs.(\ref{factH1}) - (\ref{factHn1}), and their role in the factorization method are discussed in detail in \cite{Green}. In this 
study we do not want to repeat that description. Instead, we note that from Eq.(\ref{factH1}) and explicit formulas, Eqs.(\ref{Thn}) and (\ref{Thna}), written for 
$n = 1$, one finds three following equations for the $\beta_1, \alpha_1$ and $a_1$ parameters
\begin{eqnarray}
   \beta_1 (\beta_1 - 1) = \ell (\ell + 1) \; \; \; , \; \; \;  \alpha_1 \beta_1 + \beta_1 \alpha_1 = 2 \beta_1 \alpha_1 = 2 Q \; \; \; , \; \; \; 
   a_1 = -\frac12 \alpha^{2}_1  \label{equats}
\end{eqnarray} 
From the first equation we obtain $\beta_1 = \ell + 1$. Another solution which corresponds to the $\beta_1 = -\ell$ value cannot be accepted, since it produces the 
wave function which is singular at the radial origin, i.e. at $r = 0$. Such solutions have no physical sense for the Coulomb two-body problem. By using the relation 
$\beta_1 = \ell + 1$ we determine the parameter $\alpha_1$: $\alpha_1 = \frac{Q}{\ell + 1}$. Then from Eq.(\ref{equats}) one finds that $a_1 = 
-\frac{Q^{2}}{2 (\ell + 1)^2}$. This expression for the parameter $a_1$ exactly coincides with total energy of the lowest bound state in a series of bound states 
with the angular momentum $\ell$.

Analogously, by substituting the expressions, Eqs.(\ref{Thn}) - (\ref{Thna}), into the formula for the $H_{n+1}$ Hamiltonian, Eq.(\ref{factHn1}), we obtain the 
following equations for the $\alpha_n, \beta_n, \alpha_{n+1}, \beta_{n+1}$ and $a_{n+1}$ values
\begin{eqnarray}
   \beta_{n+1} (\beta_{n+1} - 1) = \beta_{n} (\beta_{n} + 1) \; \; \; , \; \; \;  2 \alpha_{n+1} \beta_{n+1} = 2 Q = 2 \alpha_{n} \beta_{n} \; \; \; , \; \; \; 
   a_{n+1} = -\frac12 \alpha^{2}_{n+1} \; \; \;   \label{equatsn}
\end{eqnarray} 
From these equations one finds that $ \beta_{n+1} = \beta_n + 1 = \ldots = \beta_{1} + n = n + \ell + 1$, $\alpha_{n+1} = \frac{Q}{n + \ell + 1}$ and $a_{n+1} = 
-\frac12 \alpha^{2}_{n+1} = -\frac{Q^2}{2 (n + \ell + 1)^2}$. This value of $a_{n+1}$ exactly coincides with the total energy of the $n-$th excited bound state 
($E_{n+1}$) in the series of bound states with the given value of $\ell$. In other words, by using this simple method one can reproduce the bound state spectra 
for the series of bound states with arbitrary $\ell$ (angular momentum). It follows from here that the factorization method also produces the whole bound state 
spectrum of the hydrogen atom which contains the bound states with different values of angular momentum $\ell$ ($\ell \ge 0$). 

Now, let us consider the energy functional $E(\Psi) (= E_1(\Psi))$ (see, e.g., \cite{Eps}), where $\Psi = \Psi(r)$ is the trial function, and the Hamiltonian $H$ 
is represented in the form of Eq.(\ref{factH1})
\begin{eqnarray}
  E(\Psi) = \frac{\langle \Psi \mid H \mid \Psi \rangle}{\langle \Psi \mid \Psi \rangle} = \frac{\langle \Psi \mid \Theta^{\ast}_1(r) \Theta_1(r) 
  \mid \Psi \rangle}{\langle \Psi \mid \Psi \rangle} + a_1 = \frac{\langle \Theta_1(r) \Psi \mid \Theta_1(r) \Psi \rangle}{\langle \Psi \mid \Psi \rangle} 
   + a_1 \; \; \; \label{E}
\end{eqnarray} 
where $a_1$ is some negative number which is uniformly defined by $\Psi$. Since the first term in the right-hand side of this equation is always positive, 
then it follows from Eq.(\ref{E}) that $\min_{\Psi} E(\Psi) = a_1$ and such a minimum is reached on the function $\Psi$ which is defined by the equation 
$\Theta_1(r) \Psi(r) = 0$. Thus, we have found the equation which allows one to obtain the ground state wave function $\Psi_1(r)$ of an arbitrary one-electron 
atom and/or ion. At the next step we consider the subspace of functions $\Phi$ which are represented in the form $\Phi(r) = \Theta^{\ast}_1(r) \Psi(r)$, where 
the function $\Psi$ is an arbitrary radial function defined in the ${\cal L}^{2}(0 \le r < \infty)$ space. It is clear any of these functions is orthogonal to 
the ground state wave function $\Psi_1(r)$, since $\langle \Theta^{\ast}_1(r) \Psi(r) \mid \Psi_1(r) \rangle = \langle \Psi(r) \mid \Theta_1(r) \Psi_1(r) \rangle 
= 0$. This means that we are dealing with the subspace of the trial functions which are represented in the form $\Phi(r) = \Theta^{\ast}_1(r) \Psi(r)$ and all 
these functions $\Phi(r)$ are orthogonal to the ground state wave funcition $\Psi_1(r)$.  

For the $\Phi(r)$ functions we can investigate the following energy functional 
\begin{eqnarray}
  F(\Phi, \Psi) =  F(\Psi) = \frac{\langle \Phi \mid \Phi \rangle}{\langle \Psi \mid \Psi \rangle} = 
  \frac{\langle \Theta^{\ast}_1(r) \Psi \mid \Theta^{\ast}_1(r) \Psi \rangle}{\langle \Psi \mid \Psi \rangle} + a_1 \; \; \; \label{F1}
\end{eqnarray} 
By using the equality, Eq.(\ref{factHn1}), one can reduce this functional to the form
\begin{eqnarray}
 F(\Psi) = \frac{\langle \Psi \mid \Theta^{\ast}_2(r) \Theta_2(r) \mid \Psi \rangle}{\langle \Psi \mid \Psi \rangle} + a_2 = \frac{\langle \Theta_2(r) \Psi \mid 
 \Theta_2(r) \mid \Psi \rangle}{\langle \Psi \mid \Psi \rangle} + a_2  = E_2(\Psi) \; \; \; \label{E1}
\end{eqnarray}
where $E_2$ is the variational energy of the first excited state of the hydrogen atom and $a_2$ is a real negative number. It is clear that the minimum of the 
functional $F(\Psi) = E_2(\Psi)$, Eq.(\ref{E1}), equals to the $a_2$ value which coincides with the total energy of the first excited state. The corresponding 
eigenfunction is defined by the equations: $\Theta_2(r) \mid \Psi \rangle = 0$ and $\Theta_1(r) \mid \Psi \rangle \ne 0$. Note that our trial functions used in 
Eq.(\ref{E1}) are already `correct' trial functions, since they do not have any non-zero component which is proportional to the ground state wave function 
$\Psi_1$. Briefly, we can say that the minimum of the `excited' energy functional equals $a_2$, Eq.(\ref{E1}), while the corresponding wave functions are 
obtained from the equations $\Theta_2(r) \mid \Psi \rangle = 0$.

Then we can repeat this procedure by considering the non-zero functions represented in the form $\Theta^{\ast}_2(r) \Phi(r) = \Theta^{\ast}_2(r) \Theta^{\ast}_1(r) 
\Psi(r)$, where the function $\Psi(r)$ is an arbitrary, in principle, radial function defined in the ${\cal L}^{2}(0 \le r < \infty)$ space. Then, with the help of 
Eqs.(\ref{factHn1}) and (\ref{equatsn}) the whole process can be repeated as many times as needed to determine all energies of the bound states and their wave 
functions. The explicit form of the ground state wave function for one-electron atom/ion with our values of $\beta_1$ and $\alpha_1$ is $\Psi_1(r) = C r^{\ell} 
\exp(-\frac{Q}{\ell + 1} r)$ where $C$ is the normalization constant. This function is the well known exact wave function of the lowest (by the energy) state in the 
series of bound states with the given value of angular momentum $\ell$ (see, e.g., \cite{Elut}). In general, the radial wave function $\Psi_{n}$ of the $n-$excited 
state can be determined from the equation $\Theta_n(r) \Psi_n(r) = 0$. Such a wave function must be orthogonal to the corresponding radial wave functions 
$\Psi_{n-1}(r), \Psi_{n-2}(r), \ldots, \Psi_{1}(r)$ of all lower bound states. This means that we can consider the radial functions $\Psi_{n}(r), \Psi_{n-1}(r), 
\ldots, \Psi_{1}(r)$ as a `basis' in the $n-$dimensional subspace in the ${\cal L}^{2}(0 \le r < \infty)$ space of the radial functions. As is well known such a basis 
in $n-$dimensional space can be orthogonalized by using a simple procedure which described in detail in many textbooks (see, e.g., \cite{Gelf}, \cite{Halms}). After 
orthogonalization we obtain the system of unit-norm radial functions which exactly coincide with the known radial functions of the hydrogenic systems (see, e.g., 
\cite{LLQ}, \cite{Sob}).  

\section{Method of Hyperspherical Harmonics}

This and two following Sections are the central part of our study, since here we generalize the factorization method to the new level in order to include applications 
to various few- and many-electron atoms and ions, or, in other words, to many-particle Coulomb systems. Let $N_e$ be the total number of bound electrons in such an atomic 
system. The approach described in the previous Section works only for one-electron atomic systems, i.e. for $N_e = 1$. For atomic systems which contain two, three, and/or 
more bound electrons we need to develop the new approach and introduce a convenient system of new notations. First, it is clear that the total number of spatial variables 
in the case of many-electron atoms is substantially larger than three and we need to use more variables to designate all electron's spatial coordinates. This problem is 
solved below by introducing the complete set of $3 N_e$ electron hyperspherical coordinates. There are also $N_e$ electron spin coordinates which are combined in 
the total electron spin $S$ (or $S(S + 1)$ value and its $z-$projection $S_z$ (see below). Second, it is also $a$ $priori$ clear that the complete sets of conserving 
quantum numbers (or sets of `good' quantum numbers) are substantially different for one- and few-electron atomic systems. In particular, the angular momentum of any 
single bound electron $\ell_i$, where $i$ = 1, 2, $\ldots, N_e$, in many-electron atoms is not conserved. However, the vector-sum of the angular momenta of all bound 
electrons ${\bf L} = \vec{\ell}_1 + \vec{\ell}_2 + \ldots + \vec{\ell}_{N_{e}}$ is conserved. Analogously, for a single atomic electron the projection of its angular moment 
at $z-$axis, i.e. $\ell_{z_{i}} (= m_{i}$) value, is not conserved, while the sum $L_z = \ell_{1,z} + \ell_{2,z} + \ldots + \ell_{N_{e},z} = m_{1,z} + m_{2,z} + \ldots + 
m_{N_{e},z} = M$ is a conserving (or good) quantum number which is often called the magnetic quantum number $M$. In general, for an arbitrary bound state in many-electron, 
non-relativistic atom/ion one finds the following set of conserving quantum numbers $L, M$ and $\pi$, where $\pi = (-1)^{\ell_1 + \ell_2 + \ldots + \ell_{N_{e}}}$ is the 
spatial parity of the atomic wave function, or spatial parity of the bound state. In addition to these three quantum numbers in any isolated atomic system with bound 
electrons one finds the two additional quantum numbers which are always conserved: (1) the total electron spin $S$ (or $S(S + 1)$), and (2) the projection of the total 
electron spin ${\bf S}$ on the $z-$axis which is designated below as $S_z$ \cite{LLQ}. The set of these five integer and semi-integer numbers 
$\Bigl[ L, M, S, S_z, \pi \Bigr]$ uniformly defines one series of bound atomic states which is usually called the atomic term (for more details, see, e.g., \cite{Sob}).           

In atomic physics the hyperspherical coordinates were introduced by Fock in 1954 \cite{Fock} when he investigated the bound state wave function of the ground $1^{1}S-$state 
in the two-electron He atom. Later these coordinates were used in accurate computations of the different bound states of the He atom \cite{Dem}. Since the middle of 1960's 
the hyperspherical coordinates have extensively been used in nuclear and hyper-nuclear few-body problems. It was found that such coordinates are appropriate to describe 
various few-body systems which are close to their dissociation threshold(s). In 1974 Knirk \cite{Knirk} re-introduced the new set of hyperspherical coordinates in atomic and 
molecular physics. The choice of the hyperspherical coordinates in atomic problems with $N_e-$bound electrons made by Knirk was different from that used earlier by Fock 
\cite{Fock}. We have found that the definition of the hyperspherical coordinates proposed by Knirk (see Section II of his paper \cite{Knirk}) is more convenient and 
appropriate for various atomic problems. 

In this study, we shall use the same hyperspherical coordinates which exactly coincide with such coordinates defined in \cite{Knirk}. In particular, the angular (or spherical) 
coordinates of each electron are designated below as $\omega_i = (\theta_i,\phi_i)$, where $i = 1, 2, \ldots, N_e$. The radial variables of each electron $r_{i}$ are defined 
exactly as in Eq.(2.1) from \cite{Knirk} and hyper-radius $r$ coincides with the expression given in Eq.(2.3) from \cite{Knirk}. In other words, we can write for the 
Cartesian coordinates of each electron 
\begin{eqnarray}
  x_i = r_i \sin \theta_i \cos \phi_i \; \; , \; \;  y_i = r_i \sin \theta_i \sin \phi_i \; \; , \; \; z_i = r_i \cos \theta_i \; \; , \; \; \label{cart}
\end{eqnarray}
where $i = 1, 2, \ldots, N_e$, while $(x_i, y_i, z_i)$ are the Cartesian coordinates of the $i-$th electron and $r_{i} = \sqrt{x^{2}_i + y^{2}_i, + z^{2}_i}$ is the spherical
radial coordinate of this electron. It is clear that $\vec{\ell}^{2}_i \; Y_{\ell m}(\theta_j, \phi_j) = - \Delta_i Y_{\ell m}(\theta_j, \phi_j) = \delta_{ij} \ell (\ell + 1) 
Y_{\ell m}(\theta_i, \phi_i)$, where $\vec{\ell}^{2}_i$ is the square of the ordinary momentum operator of the particle $i$. 

Now, we can define the atomic hyper-radius $r = \sqrt{\sum^{N_e}_{i=1} r^{2}_{i}}$ and $(N_e - 1)$ hyperspherical angles $\eta_2, \eta_3, \ldots, \eta_{3 N_e - 1}$ which
are defined by the following relations
\begin{eqnarray}
  & & r_{N_e} = r \cos \eta_{N_e} \; \; , \; \; r_{N_e-1} = r \sin \eta_{N_e} \cos \eta_{N_e-1} \; , \;  r_{N_e-2} = r \sin \eta_{N_e} \sin \eta_{N_e-1} \cos \eta_{N_e-2} ,
  \ldots , \nonumber \\
 & & r_{2} = r \sin \eta_{N_e} \sin \eta_{N_e-1} \ldots \sin \eta_{3} \cos \eta_{2} \; \; , \; \; r_{1} =  r \sin \eta_{N_e} \sin \eta_{N_e-1} \ldots \sin \eta_{3} 
   \sin \eta_{2} \label{HHa} 
\end{eqnarray} 
The set of $(3 N_e - 1)$ angular variables (compact variables) is designated below by the letter $\Omega$, i.e. $\Omega = (\eta_{2}, \eta_{3}, \ldots, \eta_{N_e}, \omega_{1}, 
\omega_{2}, \ldots, \omega_{N_e})$. Analogously, the partial set of $(3j - 1)-$ angular variables is designated below by the letters $\Omega_j$ (= $\eta_{2}, \eta_{3}, \ldots, 
\eta_{j}, \omega_{1}, \omega_{2}, \ldots, \omega_{j})$ for $j = 2, 3, \ldots, N_e$ and $\Omega_{N_e} = \Omega$. These angular variables describe all angular configurations in 
the cluster of $j$ bound electrons. The square of the generalized angular momentum operator for the cluster of $j$ bound particles/electrons is defined by the following 
recursive relation
\begin{eqnarray} 
  \Lambda^{2}_{j}(\Omega_{j}) = -\frac{\partial^{2}}{\partial \eta^{2}_{j}} - \frac{(3 j - 4) \cos^{2} \eta_{j} - 2 \sin^{2} \eta_j}{\sin \eta_j \cos \eta_{j}} 
 \frac{\partial}{\partial \eta_j} + \frac{\Lambda^{2}_{j-1}(\Omega_{j-1})}{\sin^{2} \eta_{j}} + \frac{\vec{\ell}^{2}_j}{\cos^{2} \eta_{j}} \label{Laplj}
\end{eqnarray}
with the following `initial' condition: $\Lambda^{2}_{1}(\Omega_{1}) = \vec{\ell}^{2}_1(\omega_1)$. 

The $3 N_e$ dimensional `total' Laplacian has a very simple form in the hyperspherical coordinates 
\begin{eqnarray} 
 \nabla^2_{N_{e}} = \sum^{N_e}_{i=1} \nabla^{2}_{i=1} = \frac{\partial^2}{\partial r^2} + \frac{3 N_e - 1}{r} \frac{\partial}{\partial r} - \frac{\Lambda^2_{N_e}(\Omega)}{r^{2}}
 \label{LaplNe}
\end{eqnarray}
This term is proportional to the kinetic energy of an atom/ion which contains $N_e$ bound electrons (see below). The definition of the hyperspherical coordinates is completed 
by specifying the volume element in this coordinates $d\tau = r^{3 N_e - 1} dr d\Omega$, where $d\Omega$ is the differential surface area on the $3 N_e-$dimensional hypersphere, 
i.e.
\begin{eqnarray} 
  d\Omega = \prod^{N_e}_{j=2} (\cos^{2} \eta_j \sin^{3 j - 4} \eta_{j} d\eta_{j}) \prod^{N_e}_{i=1} (\sin \theta_i d\omega_{i}) =
  \prod^{N_e}_{j=2} (\cos^{2} \eta_j \sin^{3 j - 4} \eta_{j} d\eta_{j}) \prod^{N_e}_{i=1} (\sin \theta_i d\theta_{i} d \phi_{i}) \; \; \label{elsurf}
\end{eqnarray}
More detail description of the hyperspherical coordinates and analysis of their properties can be found, e.g., in \cite{Knirk} and in a large number of papers, books and 
textbooks on the method of hyperspherical harmonics and its applications to different problems from atomic, molecular and nuclear physics (see, e.g., \cite{Fomin} - 
\cite{Pelikan} and references therein). 

The $(2 N_e - 1)$ Laplace operators $\Lambda^{2}_{2}(\Omega_{2}), \ldots, \Lambda^{2}_{j}(\Omega_{j}), \ldots, \Lambda^{2}_{N_e}(\Omega), \vec{\ell}^{2}_1(\omega_1), \ldots,
\vec{\ell}^{2}_j(\omega_j), \ldots, \vec{\ell}^{2}_{N_{e}}(\omega_{N_e})$ depend upon different sets of angular variables. Therefore, these operators commute with each other 
and they have a common system of eigenfunctions. These eigenfunctions are represented in the form of products of eigenfunctions of the partial $(2 N_e - 1)$ Laplace operators
$\Lambda^{2}_{j}(\Omega_{j})$ and $\vec{\ell}^{2}_k(\omega_k)$. These eigenfunctions can be chosen as the `natural' basis set in the $(3 N_e - 1)$ angular (compact) space 
$\Omega$. It is clear that each of these basis functions includes the product of the spherical harmonics of each electron, i.e. ${\cal Y}(\Omega) 
\sim Y_{\ell_1 m_1}(\theta_1,\phi_1) Y_{\ell_2 m_2}(\theta_2,\phi_2) \ldots Y_{\ell_{N_e} m_{N_e}}(\theta_{N_e},\phi_{N_e})$. The eigenfunctions of the $N_e - 1$ hyperspherical 
angles $\eta_2, \eta_3, \ldots, \eta_{N_e}$ are the polynomial functions which are usually expressed in terms of the Jacobi (spherical) polynomials $P^{(\alpha,\beta)}_{n}(x)$ 
\cite{GR}, \cite{AS}. The products of eigenfunctions of all $(2 N_e - 1)$ differential operators $\Lambda^{2}_{2}(\Omega), \ldots, \Lambda^{2}_{j}(\Omega_{j}), \ldots, 
\Lambda^{2}_{N_e}(\Omega_{2}), \vec{\ell}^{2}_1(\omega_1), \ldots, \vec{\ell}^{2}_{N_e}(\omega_{N_e})$ mentioned above which depend upon the $2 N_e$ angular and $N_e - 1$ 
hyperangular variables are called the hyperspherical harmonics, or HH functions. In this study to designate the HH functions we use the notation 
${\cal Y}_{\vec{K}(b),\vec{\ell}(b),\vec{m}(b)}(\Omega)$, where $\vec{K}(b), \vec{\ell}(b), \vec{m}(b)$ is the multi-index of the hyperspherical harmonics. The numerical value 
of each component of this multi-index is uniformly related with the eigenvalue(s) of the corresponding Laplace operator mentioned above.     

In actual atomic computations only those hyperspherical harmonics (HH) are important which have the correct permutations symmetry between all bound electrons. In some earlier 
works these hyperspherical harmonics were called the `physical' (or actual) HH. For atomic systems the physical harmonics can be constructed, e.g., with the use of the 
projection operators ${\cal P}^{S S_{z}}_{L M \pi}$ for the given atomic term $\Bigl[ L, M, S, S_z, \pi \Bigr]$. The explicit construction of such projectors is well described 
in a number of original papers. For simple atomic systems, e.g., for the two-electron atoms/ions the explicit construction of such projection operators is very simple (see, 
e.g., \cite{Our1}). The physical hyperspherical harmonics are extensively used in various problems of few-body physics, including description of many different atomic systems 
(see, e.g., \cite{Avery} and references therein). 

\section{Factorization method for few- and many-electron atoms and ions}

In hyperspherical coordinates the Hamiltonian of an atom which contains $N_e$ bound electrons is written in the form \cite{Fock}, \cite{Dem} (see, also \cite{Knirk})
\begin{eqnarray}
  H(r, \Omega) =  -\frac{1}{2} \Bigl[ \frac{\partial^2}{\partial r^2} + \frac{3 N_e - 1}{r} \frac{\partial}{\partial r} - \frac{\Lambda^2_{N_e}(\Omega)}{r^{2}} \Bigr] 
  + \frac{W(\Omega)}{r} \; \; \; \label{HHH}
\end{eqnarray}
where $\Lambda^2_{N_e}(\Omega)$ is the hypermomentum of the atom, while $W(\Omega)$ is the hyperangular part of the interaction (Coulomb) potential which includes
electron-nucleus and electron-electron parts. For an atom with $N_e$ bound electrons the electron-nucleus term contains $N_e$ terms, while the electron-electron part
includes the $\frac{N_e (N_e - 1)}{2}$ terms. Now, we can consider the matrix of the operator $H(r, \Omega)$ in the basis of hyperspherical harmonics (or HH-basis, for 
short), i.e.  
\begin{eqnarray}
 \hat{H}_{ab}(r) = \langle {\cal Y}_{\vec{K(a)},\vec{\ell}(a),\vec{m}(a)}(\Omega) \mid H(r, \Omega) \mid {\cal Y}_{\vec{K}(b),\vec{\ell}(b),\vec{m}(b)}(\Omega) \rangle
  \; \; \; \label{HamlHH} 
\end{eqnarray}
where ${\cal Y}_{\vec{K}(c),\vec{\ell}(c),\vec{m}(c)}(\Omega)$ are the physical hyperspherical harmonics (see above), $\vec{K(c)} = (K_1, K_2, \ldots. K_{g_{K}}), \vec{\ell}(c) 
= (\ell_1, \ell_2, \ldots, \ell_{g_{\ell}})$ and $\vec{m}(c) = (m_1, m_2, \ldots, m_{g_{m}})$ are the multi-indeces (or vector-indeces) which uniformly define the hyperspherical 
harmonics ${\cal Y}_{\vec{K(a)},\vec{\ell}(a),\vec{m}(a)}(\Omega)$. In turn, these multi-indices of the hyperspherical harmonics are determined by the atomic state (or atomic 
term) considered in calculations. In actual computations the dimensions of these vector-indices $g_{K}, g_{\ell}$ and $g_{m}$ should be minimal, since all hyperspherical 
harmonics applied in numerical computations are the physical HH. This means that these HH have the correct permutation symmetry, or, in other words, correct symmetry in respect 
to the required permutations of all electron indices. For instance, the hyperspherical harmonics which are needed in bound state calculations of the singlet 
${}^{1}S(L = 0)-$states of the helium atom are written in the form $\mid K, \ell, \ell \rangle = \mid K, \ell \rangle$, where $K = 0, 2, 4, \ldots, 2n$ is their hypermomentum 
(index) , while $\ell \ge 0$ (and $\ell \le \frac{K}{2}$) is the second index (more details can be found in \cite{Our1}). In other words, for this atomic system each physical 
HH is designated by the two-component multi-index $(K, \ell)$, i.e. in the notations introduced above one finds $g_{K} = 1, g_{\ell} = 1$ and $m_{g_{m}} = 0$. Below, we shall 
designate the hyperspherical matrix of the Hamiltonian $\hat{H}_{ab}(r)$ by using the same notation $H$, or $H(r)$ (as we did in the second Section). It should be mentioned that 
$H(r)$ is the differential operator in respect to the hyper-radius $r$ of the second order. The explicit form of the $H(r)$ Hamiltonian operator is
\begin{eqnarray}
  H(r, \Omega) =  -\frac{1}{2} \Bigl[ \frac{\partial^2}{\partial r^2} + \frac{3 N_e - 1}{r} \frac{\partial}{\partial r} - \frac{\hat{K} (\hat{K} + 3 N_e - 2)}{r^{2}} \Bigr] 
  + \frac{\hat{W}}{r} \; \; \; \label{HHH1}
\end{eqnarray}
where $\hat{K}$ is the matrix of hypermomentum which is diagonal the basis of `physical' HH, or in $K-$representation for short. 

In our earlier study \cite{Fro1986} we have shown that the matrix of the atomic Hamiltonian $H(r)$, Eq.(\ref{HHH1}), which contains $N_e$ bound electrons is always factorized, 
i.e. it is represented in the form
\begin{eqnarray}
  H = \Theta^{\ast}_1(r) \Theta_1(r) + \hat{a}_1 \; \; \; \label{factHa}
\end{eqnarray} 
where $\hat{a}_1$ is a matrix defined below, while the operator $\Theta_1(r)$ and its adjoint operator $\Theta^{\ast}_1(r)$ are the first-order differential operators 
defined as follows
\begin{equation}
 \Theta_1(r) = \frac{1}{\sqrt{2}} \Bigl[ -\Bigl(\frac{\partial}{\partial r} + \frac{3 N_e - 1}{2 r}\Bigr) + \frac{\hat{\beta}_1}{r} + \hat{\alpha}_1 \Bigl] \; \; \; 
 \label{Thm1}
\end{equation}
and  
\begin{equation}
 \Theta^{\ast}_1(r) = \frac{1}{\sqrt{2}} \Bigl[ \Bigl(\frac{\partial}{\partial r} + \frac{3 N_e - 1}{2 r}\Bigr) + \frac{\hat{\beta}_1}{r} + \hat{\alpha}_1 \Bigl] 
  \; \; \; \label{Thm1c}
\end{equation}
where the notations $\hat{\beta}_1, \hat{\alpha}_1$ and $\hat{a}_1$ in Eqs.(\ref{factHa}) - (\ref{Thm1c})) stand for the symmetric, infinite-dimensional, in principle, matrices 
which do not commute with each other. In actual applications the dimensions of these matrices coincide with the total number of hyperspherical harmonics used. By substituting 
these two expressions, Eqs.(\ref{Thm1}) - (\ref{Thm1c}), into Eq.(\ref{factHa}) one finds the three following equations for the $\hat{\alpha}_1, \hat{\beta}_1$ and 
$\hat{a}_1$ matrices:
\begin{eqnarray}
    & & \hat{\beta}_1 (\hat{\beta}_1 - 1) = \Bigr(\hat{K} + \frac{3 N_e - 1}{2}\Bigl) \Bigr(\hat{K} + \frac{3 N_e - 1}{2} - 1\Bigl)  \; \; \; \label{eqx} \\
    & & \hat{\alpha}_1 \hat{\beta}_1 + \hat{\beta}_1 \hat{\alpha}_1 = 2 \hat{W} \; \; \; \label{eqy} \\
    & & \hat{a}_1 = -\frac12 \hat{\alpha}^2_1 \; \; \; \label{eqz}
\end{eqnarray}
where the matrix of hypermomentum $\hat{K}$ is a diagonal matrix in the basis of hyperspherical harmonics (or, in $K-$representation, for short). Solution of Eq.(\ref{eqx}) 
is written in the form 
\begin{equation}
    \hat{\beta}_1 = \hat{K} + \frac{3 N_e - 1}{2}   \; \; \; \label{eqxx}
\end{equation}
where we use the fact that the atomic wave function is regular at $r = 0$, or at the atomic nucleus. As follows from this equation the matrix $\hat{\beta}_1$ is diagonal in 
$K-$representation.  Below, we apply only this $K-$representation, since it substantially simplifies a large number of formulas derived below. In particular, by using 
Eq.(\ref{eqy}) and the formula from \cite{Belman} (see Chapter 10, \$ 18) we can write the explicit expression for the $\hat{\alpha}_1$ matrix
\begin{equation}
    \hat{\alpha}_1 = 2 \int_{0}^{+\infty} \exp(-\hat{\beta}_1 t) \hat{W} \exp(-\hat{\beta}_1 t) dt  \; \; \; \label{eqyy}
\end{equation}
Since the $\hat{\beta}_1$ matrix is diagonal, then for the $(ij)-$matrix element of the $\hat{\alpha}_1$ matrix takes the form 
\begin{equation}
  \Bigl[ \hat{\alpha}_1 \Bigr]_{ij} = \frac{2 W_{ij}}{[\beta_1]_{ii} + [\beta_1]_{jj}} = \frac{2 W_{ij}}{[\beta_{1}]_{i} + [\beta_{1}]_{j}} = 
   \frac{2 W_{ij}}{K_{i} + K_{j} + 3 N_e - 1}
\end{equation}
Finally, we can determine the $\hat{a}_1$ matrix from Eq.(\ref{eqz}). In particular, for the $(ij)-$matrix elements of the $\hat{a}_1$ matrix one finds
\begin{equation}
  \Bigl[ \hat{a}_1 \Bigr]_{ij} = - 2 \sum_{k} \frac{W_{ik}}{\beta_{i} + \beta_{k}} \cdot \frac{W_{kj}}{\beta_{k} + \beta_{j}} = 
     - 2 \sum_{k} \frac{1}{\beta_{i} + \beta_{k}} \Bigl[ W_{ik} W_{kj} \Bigr] \frac{1}{\beta_{k} + \beta_{j}}  \; \; \; \label{eqzz}
\end{equation}

At the second stage of the procedure, we introduce the set of radial operators $\Theta_n(r)$, where $n = 2, 3, \ldots$, which are similar to the operator $\Theta_1(r)$ defined 
above (see, Eq.(\ref{Thm1})), i.e.
\begin{equation}
 \Theta_n(r) = \frac{1}{\sqrt{2}} \Bigl[ -\Bigl(\frac{\partial}{\partial r} + \frac{3 N_e - 1}{2 r}\Bigr) + \frac{\hat{\beta}_n}{r} + \hat{\alpha}_n \Bigl] \; \; \; 
 \label{Thmn}
\end{equation}
Therefore, its adjoint operator takes the form 
\begin{equation}
 \Theta^{\ast}_n(r) = \frac{1}{\sqrt{2}} \Bigl[ \Bigl(\frac{\partial}{\partial r} + \frac{3 N_e - 1}{2 r}\Bigr) + \frac{\hat{\beta}_n}{r} + \hat{\alpha}_n \Bigl] 
  \; \; \; \label{Thmnc}
\end{equation}
In order to construct the correct and logically closed algorithm of the factorization method the following conditions must be obeyed 
\begin{eqnarray}
  \Theta_n(r) \Theta^{\ast}_n(r) + \hat{a}_n = H_{n+1} = \Theta^{\ast}_{n+1}(r) \Theta_{n+1}(r) + \hat{a}_{n+1} \; \; \; \label{factHmn}
\end{eqnarray}
for $n = 1, 2, \ldots$. By substituting the explicit expressions, Eqs.(\ref{Thmn}) and (\ref{Thmnc}) into Eq.(\ref{factHmn}) we obtain the following equations for the 
$\hat{\beta}_n, \hat{\beta}_{n+1}, \hat{\alpha}_{n}, \hat{\alpha}_{n+1}, \hat{a}_n$ and $\hat{a}_{n+1}$ matrices
\begin{eqnarray}
 & &   \hat{\beta}_{n+1} (\hat{\beta}_{n+1} - 1) = \hat{\beta}_{n} (\hat{\beta}_{n} + 1) \; \; \; , \; \; \; \label{I1} \\ 
 & & \hat{\alpha}_{n} \beta_{n} + \beta_{n} \hat{\alpha}_{n} = 2 \hat{W} = \alpha_{n+1} \beta_{n+1} + \beta_{n+1} \hat{\alpha}_{n+1} \; \; \; , \; \; \; \label{I2} \\
 & & \hat{a}_{n} = -\frac12 \alpha^{2}_{n} \; \; \; , \; \; \; \hat{a}_{n+1} = -\frac12 \alpha^{2}_{n+1} \; \; \;   \label{I3}
\end{eqnarray} 
These matrix equations look very similar to the analogous numerical equations mentioned in Section II (see, Eqs.(\ref{equatsn})). However, these equations Eqs.(\ref{I1}) 
- (\ref{I3}), are written for the symmetric, infinite-dimensional matrices, which do not commute with each other, e.g., the $\hat{\beta}_n$ matrix do not commute with the 
$\hat{\alpha}_{n}$ and $\hat{a}_{n+1}$ matrices, etc. Solution of these equations, Eqs.(\ref{I1}) - (\ref{I3}), regular at $r = 0$ is written in the following form(s)
\begin{eqnarray}
 & & \hat{\beta}_{n+1} = \hat{\beta}_{n} + 1 = \ldots = \hat{\beta}_1 + n = \hat{K} + \frac{3 N_e - 1}{2} + n \; \; \; \label{eqxxm} \\
 & & \hat{\alpha}_{n+1} = 2 \int_{0}^{+\infty} \exp(-\hat{\beta}_{n+1} t) \hat{W} \exp(-\hat{\beta}_{n+1} t) dt  \; \; \; \label{eqyym} \\
 & & \hat{a}_{n+1} = -\frac12 \alpha^{2}_{n+1} \; \; \; \label{eqzzm}
\end{eqnarray}

The second equaition, Eq.(\ref{eqyym}), produces the following explicit expression for the $(ij)-$matrix element of the $\hat{\alpha}_{n+1}$ matrix
\begin{equation}
  \Bigl[ \hat{\alpha}_{n+1} \Bigr]_{ij} = \frac{2 W_{ij}}{[\beta_{n+1}]_{ii} + [\beta_{n+1}]_{jj}} = \frac{2 W_{ij}}{[\beta_{1}]_{i} + [\beta_{1}]_{j} + 2 n} 
   = \frac{2 W_{ij}}{K_{i} + K_{j} + 3 N_e - 1 + 2 n} \; \; \; \label{fres1}
\end{equation}
where $[\beta_{1}]_{i}$ is the diagonal $(ii)-$matrix element of the diagonal $\hat{\beta}_{1}$ matrix, i.e. $[\beta_{n+1}]_{ij} = \delta_{ij} [\beta_{n+1}]_{ii} =
\delta_{ij} [\beta_{n+1}]_{i}$ and $[\beta_{1}]_{ij} = \delta_{ij} [\beta_{1}]_{ii} = \delta_{ij} [\beta_{1}]_{i}$. This leads to the following analytical formula 
for the $(ij)-$matrix elements of the $\hat{a}_{n+1}$ matrix
\begin{eqnarray} 
 & & \Bigl[ \hat{a}_{n+1} \Bigr]_{ij} = - 2 \sum_{k} \frac{W_{ik}}{[\beta_{1}]_{i} + [\beta_{1}]_{k} + 2 n} \cdot \frac{W_{kj}}{[\beta_1]_{k} + [\beta_1]_{j} + 2 n} 
  \nonumber \\
 &=& - \frac12 \sum_{k} \frac{1}{b_{ik} + n} \Bigl[ W_{ik} W_{kj} \Bigr] \frac{1}{b_{kj} + n} \; \; \; \label{eqzzm1} \\
 &=& - 2 \sum_{k} \frac{1}{K_{i} + K_{k} + 2 n + 3 N_e - 1} \Bigl[ W_{ik} W_{kj} \Bigr] \frac{1}{K_{k} + K_{j} + 2 n + 3 N_e - 1} \nonumber
\end{eqnarray}
where $b_{ik} = \frac12 ([\beta_{1}]_{i} + [\beta_{1}]_{k})$ and  $b_{kj} = \frac12 ([\beta_{1}]_{k} + [\beta_{1}]_{j})$, while $K_{i}$ are the matrix elements of the 
diagonal $\hat{K}$-matrix (the matrix of hypermomentum) and $n \ge 0$, where $n$ is the radial quantum number (integer, non-negative). Formally, the formula, 
Eq.(\ref{eqzzm1}), is a direct generalization of the Bohr's formula, originally derived by N. Bohr (in 1913) for the hydrogen atom, to an atom/ion which contains $N_e$ 
bound electrons. In Quantum Mechanics the same formula for the spectra of the hydrogen atom was derived by W. Pauli in 1926 \cite{Pauli}. For $N_e = 1$ the formula 
Eq.(\ref{eqzzm1}) exactly coincides with the formula Eq.(\ref{spectr1}) (in atomic units). Indeed, in this case $3 N_e - 1 = 2$, $\hat{W}_{ij} = -Q \delta_{ij}, K_i = K_j
= \ell$ and $\ell$ is the good quantum number. Therefore, one finds from Eq.(\ref{eqzzm1}) $E_i = \Bigl[ \hat{a}_{n+1} \Bigr]_{ii} = - \frac{Q^2}{2 (\ell + 1 + n)^2}$.
For few- and many-electron atoms the situation is more complicated, since for such systems we need to know the explicit forms of the radial part of the total wave 
functions. This problem is discussed in the next Section.
 
\section{Bound state wave functions}

Let us discuss an approach which can be used to determine the wave functions of the bound states in atoms/ions which contain $N_e$ ($N_e \ge 1$) bound electrons. This 
approach is based on the basic equations of matrix mechanics derived above and has a number of similarities with the analogous method used in Section II for one-electron 
atoms and ions. In particular, the ground (bound) state wave functions can be determined from the differential equation of the first order $\Theta_1(r) \Psi(r) = 0$. The 
explicit form of this equation is
\begin{eqnarray}
 \Bigl[ -\Bigl(\frac{\partial}{\partial r} + \frac{3 N_e - 1}{2 r}\Bigr) + \frac{\hat{\beta}_1}{r} + \hat{\alpha}_1 \Bigl] \Psi(r) = 
 \Bigl[ -\frac{\partial}{\partial r} + \frac{\hat{K}}{r} + \hat{\alpha}_1 \Bigl] \Psi(r) = 0 \; \; \; \label{wf1}
\end{eqnarray}
where $\hat{K}$ is the diagonal matrix of the hypermomentum, i.e. $\hat{K} = \hat{\beta}_1 - \frac{3 N_e + 1}{2}$. To solve this equation we can represent the function 
$\Psi(r)$ in the form $\Psi(r) = r^{\hat{K}} \exp(\lambda r) \bf{C}$, where $\lambda$ is a real (always negative) numerical constant defined below and $\bf{C}$ is the 
numerical vector, i.e. each component of this vector does not depend upon the hyper-radius $r$. Substitution of the function $\Psi(r)$ in this form into Eq.(\ref{wf1}) 
reduces this equation to the form
\begin{eqnarray} 
 \Bigl[ - \frac{\hat{K}}{r} - \lambda + \frac{\hat{K}}{r} + \hat{\alpha}_1 \Bigl] \Psi(r) = \Bigl[ \hat{\alpha}_1 - \lambda \Bigl] r^{\hat{K}} \exp(\lambda r) \bf{C} = 0 
  \; \; \; \label{wf151}
\end{eqnarray}
In other words, to determine the numerical value of $\lambda$ ($\lambda < 0$ or $\lambda = - \mid \lambda \mid$) we need to solve the following generalized eigenvalue 
problem: $(\hat{A} - \lambda \hat{B}) \bf{C} = 0$, where the matrix elements of the $\hat{A}$ and $\hat{B}$ matrices are defined by the following equations
\begin{eqnarray}
 & & [\hat{A}]_{ij} = \frac{1}{{\cal N}_{i} {\cal N}_{j}} \int_{0}^{+\infty} r^{K_{i}} \Bigr( \hat{\alpha}_1 \Bigl)_{ij} r^{K_{j}} \exp(2 \lambda r) r^{3 N_{e} - 1} dr 
  = \frac{\Gamma(K_i + K_j + 3 N_e)}{{\cal N}_{i} {\cal N}_{j} (2 \mid \lambda \mid)^{K_i + K_j + 3 N_e}} \Bigr( \hat{\alpha}_1 \Bigl)_{ij} \label{wf152a} \\
  & & [\hat{B}]_{ij} = \frac{\delta_{ij}}{{\cal N}^{2}_{i}} \int_{0}^{+\infty} r^{2 K_{i}} \exp(2 \lambda r) r^{3 N_{e} - 1} dr = 
   \frac{\Gamma(2 K_i + 3 N_e)}{{\cal N}^2_{i} (2 \mid \lambda \mid)^{2 K_i + 3 N_e}} \label{wf152b}   
\end{eqnarray}
where $\Gamma(x)$ is the Euler $\Gamma-$function. In actual applications the numbers $K_i, K_j$ and $N_e$ are integer and we can apply the formula $\Gamma(a) = (a - 1)!$
As follows from the second equations the matrix $\hat{B}$ is diagonal and all its eigenvalues (i.e. diagonal elements) are positive. Now, we can choose the normalization 
constant ${\cal N}_{i}$ in Eq.(\ref{wf152b}) as follows 
\begin{eqnarray}
    {\cal N}_i = \sqrt{\frac{(2 \mid \lambda \mid)^{2 K_i + 3 N_e}}{\Gamma(2 K_i + 3 N_e)}} \; \; \; \label{wf152c} 
\end{eqnarray}
This transforms the the matrix $\hat{B}$ into the unit matrix. The original problem is reduced to the regular eigenvalue problem, i.e. to the equation $(\hat{\tilde{A}} - 
\lambda) \bf{C} = 0$, where $\hat{\tilde{A}}$ is the $\hat{A}$ matrix in the new `normalized' basis, i.e. 
\begin{eqnarray}
  [\hat{\tilde{A}}]_{ij} = \frac{\Gamma(K_i + K_j + 3 N_e)}{\sqrt{\Gamma(2 K_i + 3 N_e) \Gamma(2 K_j + 3 N_e)}} \Bigr( \hat{\alpha}_1 \Bigl)_{ij} \; \; \; \label{wf153a} 
\end{eqnarray}
Note that the matrix elements of this matrix ($\hat{\tilde{A}}$) do not depend (explicitly) upon $\lambda$. If we know the numerical value of $\lambda$, then the total energy 
of the lowest energy state in the term is $E_1 = -\frac{1}{2} \lambda^2$. 

The eigenvalue problem discussed above is equivalent to the finding of the absolute minimum of the following energy functional $E_1(\Psi)$
\begin{eqnarray}
  E_1(\Psi) &=& \frac{\langle \Psi \mid H \mid \Psi \rangle}{\langle \Psi \mid \Psi \rangle} = \frac{\langle \Psi \mid \Theta^{\ast}_1(r) \Theta_1(r) 
  \mid \Psi \rangle}{\langle \Psi \mid \Psi \rangle} + \frac{\langle \Psi \mid  \hat{a}_1 \mid \Psi \rangle}{\langle \Psi \mid \Psi \rangle} \nonumber \\
  &=& \frac{\langle \Theta_1(r) \Psi \mid \Theta_1(r) \Psi \rangle}{\langle \Psi \mid \Psi \rangle} + 
  \frac{\langle \Psi \mid \hat{a}_1 \mid \Psi \rangle}{\langle \Psi \mid \Psi \rangle} \; \; \; \label{EN-e}
\end{eqnarray} 
In our strategy of minimization the trial wave functions is represented in the form of the linear combinations: $\Psi(r) = r^{\hat{K}} \exp(\lambda r) {\bf C} = \sum^{N}_{n=1} 
C_{n} r^{\hat{K}_n} \exp(\lambda r) = {\bf C} r^{\hat{K}_n} \exp(\lambda r)$. The first term in the right-hand side of Eq.(\ref{EN-e}) is always non-negative. For our choice of 
the radial trial functions this term equals zero identically. The second term in the right-hand side of Eq.(\ref{EN-e}) is always negative. Optimization of the linear 
coefficients $C_k$ in our wave function at the second stage of the method means that we are trying to make the second term in Eq.(\ref{EN-e}) as negative as possible. 

The wave functions of the excited state in the atoms/ions with the $N_e$ bound electrons are determined analogously. Let us briefly describe this process by omitting 
some obvious details. The equation which determines the wave function of the $n-$th excited states ($\Psi_n(r)$) takes the form
\begin{eqnarray} 
 \Theta_n(r) \Psi_n(r) = \Bigl[ -\frac{\partial}{\partial r} + \frac{\hat{K} + n}{r} + \hat{\alpha}_1 \Bigl] \Psi_n(r) = 0 \; \; \; \label{wf153}
\end{eqnarray}
To solve this equation we represent the wave function $\Psi_n(r)$ in the form $\Psi_n(r) = r^{\hat{K} + n} \exp(\lambda_n r) \bf{C}_n$, where $\lambda_n$ is a real (and negative) 
numerical constant defined below and ${\bf C}_n$ is the $r-$independent constant vector. By subsituting 
\begin{eqnarray} 
 \Bigl[ - \frac{\hat{K} + n}{r} - \lambda_n + \frac{\hat{K} + n}{r} + \hat{\alpha}_n \Bigl] \Psi_n(r) = \Bigl[ \hat{\alpha}_n - \lambda_n \Bigl] r^{\hat{K} + n} \exp(\lambda_n r) 
 {\bf C}_n = 0 \; \; \; \label{wf154}
\end{eqnarray}
This problem is reduced to the solution of the following generalized eigenvalue problem: $(\hat{A}_n - \lambda \hat{B}_n) {\bf C}_n = 0$, where the matrix elements of the 
$\hat{A}_n$ and $\hat{B}_n$ matrices are defined by the following equations
\begin{eqnarray}
 & & [\hat{A}_n]_{ij} = \frac{1}{{\cal N}_{i} {\cal N}_{j}} \int_{0}^{+\infty} r^{K_{i} + n} \Bigr( \hat{\alpha}_n \Bigl)_{ij} r^{K_{j} + n} \exp(2 \lambda_n r) r^{3 N_{e} - 1} 
 dr \; \; \label{wf155a} \\
  & & [\hat{B}_n]_{ij} = \frac{\delta_{ij}}{{\cal N}^{2}_{i}} \int_{0}^{+\infty} r^{2 K_{i} + 2 n} \exp(2 \lambda_n r) r^{3 N_{e} - 1} dr \; \; \label{wf155b}   
\end{eqnarray}
where the matrix $\hat{B}_n$ is diagonal and all its eigenvalues (i.e. diagonal elements) are positive. Again, we can choose the normalization constants ${\cal N}_{i}$ in the form 
\begin{eqnarray}
    {\cal N}_i = \sqrt{\frac{(2 \mid \lambda \mid)^{2 K_i + 2 n + 3 N_e}}{\Gamma(2 K_i + 2 n + 3 N_e)}} \; \; \; \label{wf155c} 
\end{eqnarray}
In this case the matrix $\hat{B}_n$ will coincide with the unit matrix. This reduces the original problem to the regular eigenvalue problem, i.e. $(\hat{{\cal A}}(n) - \lambda_n) 
{\bf C}_n = 0$, where $\lambda_{n}$ is the lowest eigenvalue of the $\hat{{\cal A}}(n)$ matrix which is the matrix $\hat{A}_n$ in the new `normalized' basis, i.e.
\begin{eqnarray}
  [\hat{{\cal A}}(n)]_{ij} = \frac{\Gamma(K_i + K_j + 2 n + 3 N_e)}{\sqrt{\Gamma(2 K_i + 2 n + 3 N_e) \Gamma(2 K_j + 2 n + 3 N_e)}} \Bigr( \hat{\alpha}_n \Bigl)_{ij} \; \; \; 
 \label{wf155f} 
\end{eqnarray}
where $n \ge 1$. Again, we note that the matrix elements of this matrix ($\hat{{\cal A}}(n)$) do not depend (explicitly) upon $\lambda$. The known numerical value of $\lambda_n$ 
determines the total energy of the $n-$th excited bound state in the same atomic term term: $E_{n+1} = -\frac{1}{2} \lambda^{2}_{n}$. Thus, by using the method of matrix 
factorization developed in this study one can find all bound states in one atomic term and their wave functions. Such wave functions must be orthogonalized to each other to form
a set of actual wave functions. Note also that the marices $\hat{\alpha}_n$ and $\hat{\alpha}_{n+1}$ which are used in this method are closely related to each other, Indeed, the 
$\hat{\alpha}_{n+1}$ matrix easily obtained from the `previous' $\hat{\alpha}_{n}$ matrix by adding the term +1 in its denominator. i.e. by the replacement $n \rightarrow n + 1$.

Furthermore, as follows from Eq.(\ref{wf155f}) for any given bound state in many-electron atoms the radial quantum number $n$ is a conserving quantum number which can be used to 
number (or locate) this bound state inside of one series of bound states which have the same values of $L, L_z (or M), S, S_z$ and $\pi$. In other words, this radial quantum 
number $n$ (or excitation index) can be used to number the bound states inside of one atomic term. In general, any bound state in the atomic term can be designated by the notation 
$\mid n, \Bigl[ L, M, S, S_z, \pi \Bigr] \rangle$, where the internal notation $\Bigl[ L, M, S, S_z, \pi \Bigr]$ designates the corresponding atomic term and $n$ is the number of 
this (bound) state in this atomic term, or, in other words, the number of excitation(s). The same notation $\mid n, \Bigl[ L, M, S, S_z, \pi \Bigr] \rangle$ can be used to designate 
the corresponding wave function(s). This `conservation' of the `radial' quantum number $n$ (in our current notation) allows one to designate the bound states in few- and 
many-electron atoms. For instance, the ground bound (doublet) state in the lithium atom can be designated by the notation $1^{2}S$-state (instead of the $2^{2}S$ notation used 
currently).

\section{Applications and Conclusions}

We have developed the method of matrix factorization which can be applied to many-electron (or many-particle) atoms, ions and molecules. Formally, this method can be used for 
arbitrary many-body systems where each pair of particles interacts by the Coulomb potential. Briefly, for each of these systems the corresponding Hamiltonian written in the 
hyperspherical multi-dimensional coordinates must be similar to the form of Eq.(\ref{HHH}) (\cite{Fock} - \cite{Knirk}). The main difference between the matrix factorization and 
`regular' (or numerical) factorization follows from the fact that in the method of matrix factorization we use a number of infinite-dimensional matrices which do not commute with 
each other. This fact complicates the procedure of matrix factorization and its applications to many-electron atomic systems. Nevertheless, we could develop the logically closed 
algorithm of the matrix factorization, and now this method can be applied to determine the bound states in a large number of actual (i.e. few- and many-electron) atomic systems. 

At the first step of the procedure we need to calculate the (symmetric) matrix of the potential energy $\hat{W}$ in the basis of the physical hyperspherical harmonics constructed 
for some atomic term $\Bigl[ L, M, S, S_z, \pi \Bigr]$. By using  this matrix it is easy to construct an infinite, in principle, consequence of matrices $\hat{{\cal A}}(n)$ each 
of which has the following matrix elements
\begin{eqnarray}
  [\hat{{\cal A}}(n)]_{ij} &=& \frac{\Gamma(K_i + K_j + 2 n + 3 N_e)}{\sqrt{\Gamma(2 K_i + 2 n + 3 N_e) \Gamma(2 K_j + 2 n + 3 N_e)}} \cdot 
  \frac{2 W_{ij}}{K_{i} + K_{j} + 3 N_e - 1 + 2 n} \; \; \label{wf158a} \\
 &=& \frac{(K_i + K_j + 2 n + 3 N_e - 1)!}{\sqrt{(2 K_i + 2 n + 3 N_e - 1)! (2 K_j + 2 n + 3 N_e - 1)!}} \cdot \frac{2 W_{ij}}{K_{i} + K_{j} + 3 N_e - 1 + 2 n} \nonumber  
\end{eqnarray}
where $n = 0, 1, 2, \ldots$. The matrix $[\hat{{\cal A}}(n)]$ is symmetric and all its eigenvalues are negative. At the second stage of the procedure we determine the lowest 
eigenvalue $\lambda_{n+1}$ of each of these matrices $\hat{{\cal A}}(n)$, where $n = 0, 1, 2, \ldots$. The total energies $E_{n+1}$ of the corresponding bound states in the atom/ion 
with $N_e$ bound electrons are simply related with the $\lambda_{n+1}$ eigenvalues by the formula $E_{n+1} = -\frac12 \lambda^{2}_{n+1}$. This gives us the complete energy spectrum 
of bound state for this atomic term $\Bigl[ L, M, S, S_z, \pi \Bigr]$. To find the corresponding wave functions one needs to use the procedure described in the previous Section 
which must include the orthogonalization of the set of wave functions with different $n$ at the final step. Briefly, to obtain the total energies of all bound states in one atomic 
term in our method we need to determine the lowest eigenvalue for each of the matrices which are included in the following (infinite) consequence of closely related matrices 
$\hat{{\cal A}}(1), \hat{{\cal A}}(2), \ldots, \hat{{\cal A}}(n), \hat{{\cal A}}(n+1), \ldots$. The matrix $\hat{{\cal A}}(n+1)$ is obtained from the matrix $\hat{{\cal A}}(n)$ by 
replacing the radial quantum number $n$ in Eq.(\ref{wf158a}) by the `next' $n + 1$ value. 

For one-electron atomic systems when $N_e = 1$ we have in Eq.(\ref{wf158a}) $K_i = K_j = \ell, W_{ij} = - Q \delta_{ij}$ and $\ell$ is the conserving quantum number of atomic angular 
moment. This leads to the answer known for the hydrogen-like atom/ions discussed above. An additional interesting fact follows directly from Eq.(\ref{wf158a}) where each term in the 
right-hand side depends upon the $n + \ell$ sum only (not, e.g., upon the $n + 2 \ell$ and/or $n + \frac13 \ell$ sums). This fact is closely related to an additional symmetry of the 
bound states in one-electron atom/ions, since we can replace the conserving quantum number $n$ (or $n_r$ in usual notation) by the $\ell$ quantum number and vice versa. The total 
energy will not change during such substitutions. For atoms with $N_e \ge 2$ such a replacement has no sense, since the hyper-radial quantum number $n$ (or $n_r$) is a conserving 
quantum number, while an arbitrary component of the diagonal matrix of hypermomentum ($K_i$ and $K_j$) does not conserve.

The method of matrix factorization developed in this study has been applied to the variational bound state calculations of bound states in the ${}^{1}S$ atomic term of the helium 
atom. In our calculations we have used 576 hyperspherical harmonics (HH). In particular, we have used all hyperspherical harmonics from the families of HH up to $K_{max} = 40$, 
some selected HH from families from $K_{max} = 44$ up to $K_{max} = 52$ and only main HH from the families of HH from $K_{max} = 56$ up to $K_{max} = 100$. The physical sense of the 
main hyperspherical harmonics was explained in detail in \cite{Fro1986}. The main hyperspherical harmonics for the bound states in the ${}^{1}S-$term of the He atom have the from 
$\mid K, \ell \rangle = \mid 4 k, 0 \rangle$, where $k$ is any non-negative integer number. The total energies of some lower-lying bound states in the ${}^{1}S-$term in the He atom 
obtained by using our method of matrix factorization are $E_1$ = -2.9037175 $a.u.$,  $E_2$ = -2.144954 $a.u.$, $E_3$ = -2.06033 $a.u.$, $E_4$ = -2.0318 $a.u.$ The `exact' total 
energies obtained in our earlier calculations for these bound states are \cite{FroHe} $E_1$ = -2.903724377034119598311159245194405(5) $a.u.$, $E_2$ = -2.145974046054417415(10) $a.u.$, 
$E_3$ = -2.06127198974090848(5) $a.u.$, $E_4$ = -2.03358671703072520(7) $a.u.$ These values are significantly more accurate than the total energies found with the use of our procedure 
based on the hyperspherical harmonics. This can be explained by the known fact (see, e.g., \cite{Fro1986}) that hyperspherical expansion is not very effective approach to describe 
electron-electron correlations in actual atoms and ions. However, the overall accuracy of the method based on the hyperspherical expansion can be increased drastically, e.g., by 
increasing the total number of the main HH used and/or by considering the coherent hyperspherical states.    
 
The method of matrix factorization allows one to determine the bound state spectra of many-electron (but non-relativistic!) atoms, ions and molecules. This means that by using 
our method one can determine, in principle, all bound state energies and corresponding wave functions. At the following stages these wave functions can be applied to evaluate 
various bound state properties, including lowest-order relativistic and QED corrections for different atoms, ions and molecules. Formally, the method of matrix factorization 
allows one to obtain analytical and semi-analytical answers to numerous questions about atomic structure of the few- and many-electron (non-relativistic) atoms, ions and light 
molecules. In many cases, however, the obtained answers and solutions are often written in the matrix form which is directly related to the original matrix form of the matrix 
(quantum) mechanics.  

It should be mentioned that since Niels Bohr published (in 1912) his famous formula for the energy levels in the hydrogen atom a large number of people have tried to derive analogous 
formulas for few- and many-particle atoms and ions. In some studies it was assumed that all bound state of an atom can be found as the roots of some polynomial/analytical function, 
while another direction was based on analytical/numerical diagonalization of some `universal' matrix. All these attempts have failed. Equations derived in this study explain the 
reasons of such a failure. Note also that our method of matrix factorization of the Coulomb many-particle Hamiltonians has been developed with a substantial time delay. The basic 
equations of the method of matrix factorization have been produced at the end of 1978 when I was a student. Nevertheless, the complete version of the method has been formulated only 
in April this year. Unfortunately, this paper cannot be published in the middle of 1950's, or even earlier, when Dirac, Fock and Heisenberg were around. At the same time a large 
number of competing computational methods have extensively been developed and applied to atomic physics. Some of these methods became very effective, relatively simple and fast 
procedures. However, even now the method of matrix factorization has a great potential for future development and various modifications, since it is based on the internal `ladder' 
structure of the Coulomb Hamiltonians. Furthermore, the matrix factorization is the new, relatively simple and advanced approach which can be used to investigate the few- and 
many-body Coulomb problems and determine the bound states in such systems. In particular, our method can be used to understand some interesting details of atomic spectra and 
substantially simplify accurate bound state computations of different systems known in atomic and molecular physics. 

Finally, we want to emphasize that the method of matrix factorization is substantially based on the ladder structure of the Hamiltonians of the Coulomb many-body systems. In this
study we discovered the method which uses this ladder structure of the Coulomb Hamiltonians and allows one to determine all bound states in any few- and/or many-body Coulomb 
system. Based on the ladder structure of the Coulomb Hamiltonians we can predict that this method can be used as a very effective tool for theoretical and numerical investigation 
of the bound state spectra in all Coulomb atomic and molecular systems. For instance, the method of matrix factorization allows one to study general dependencies of the total 
energies of different bound states in the few- and many-electron atoms/ions upon good quantum numbers $a$ $priory$ known for such quantum systems. Note also that for Coulomb 
three-body systems we have developed another method \cite{Fro02} which is also based on the ladder structure of the Hamiltonians, but allows one to obtain the corresponding 
eigenvalues (and eigenfunctions) to substantially better accuracy.

\end{document}